[1994/12/01]
\documentclass{aomart}
\pdfoutput=1
\usepackage{tikz}
\usetikzlibrary{calc,arrows,automata}
\usetikzlibrary{matrix,positioning,arrows,decorations.pathmorphing,shapes}
\usetikzlibrary{shapes,snakes}

\usepackage{algorithm}
\usepackage{algorithmic}

\usepackage{amssymb}
\usepackage{amsthm}
\usepackage{amsmath}

\usetikzlibrary{matrix,positioning,arrows,decorations.pathmorphing,shapes}
\usepackage{fancyhdr}

\chardef\bslash=`\\ 

\newtheorem[{}\it]{thm}{Theorem}[section]

\newtheorem{theorem}[thm]{Theorem}

\theoremstyle{definition}

\newtheorem*[{}\it]{notation}{Notation}

\newcommand{\eval}[2][\right]{\relax
  \ifx#1\right\relax \left.\fi#2#1\rvert}

\title[A Full Characterization of Irrelevant Components in Diameter Constrained Reliability]
{A Full Characterization of Irrelevant Components in Diameter Constrained Reliability}

\author{Eduardo Canale, Pablo Romero, Gerardo Rubino}
\email{canale@fing.edu.uy; pablo.promero@inria.fr; rubino@inria.fr} 
\address{Facultad Polit\'ecnica, Universidad Nacional de Asunci\'on.\\ 
Ruta Mcal. Estigarribia, Km. 10,5. San Lorenzo - Paraguay.\\
Inria Rennes, Bretagne-Atlantique, Campus de Beaulieu.\\
PC 35042, RENNES Cedex}

\subject{primary}{matsc2000}{123456789}
\subject{secondary}{matsc2000}{FFFF}

\volumenumber{1}
\issuenumber{1}
\publicationyear{2014}

\begin{document}
 
\begin{abstract}
In classical network reliability analysis, the system under study is a network with perfect nodes but imperfect link, that 
fail stochastically and independently. There, the goal is to find the probability that the resulting random graph is connected, 
called \emph{reliability}. 
Although the exact reliability computation belongs to the class of $\mathcal{NP}$-Hard problems, the literature offers three 
exact methods for exact reliability computation, to know, Sum of Disjoint Products (SDPs), Inclusion-Exclusion and Factorization.\\

Inspired in delay-sensitive applications in telecommunications, H\'ector Cancela and Louis Petingi defined in 2001 the diameter-constrained 
reliability, where terminals are required to be connected by $d$ hops or less, being $d$ a positive integer, called diameter. \\

Factorization theory in classical network reliability is a mature area. However, an extension to the diameter-constrained context 
requires at least the recognition of irrelevant links, and an extension of deletion-contraction formula. 
In this paper, we fully characterize the determination of irrelevant links. Diameter-constrained reliability invariants are presented, 
which, together with the recognition of irrelevant links, represent the building-blocks for a new factorization theory. 
The paper is closed with a discussion of trends for future work. 
\end{abstract}

\maketitle
\tableofcontents


\section{Motivation}
The diameter-constrained reliability measure was introduced in 2001 by H\'ector Cancela and Louis Petingi, inspired in delay sensitive applications~\cite{PR01}.  
In telecommunications, there are several problems where the diameter (or the number of hops) in the communication is a major cause of concern. 
In flooding-based systems, the number of hops should be controlled in order to avoid network congestion. Peer-to-peer networks 
originally support file discovery protocols by means of flooding~\cite{Adar00freeriding}.   
Internet Protocol version 6 (IPv6) has a ``Hop limit'' field, reserved for these cases~\cite{ietf:ipv6}. 
Another hot topic in network design is fiber optics deployment. There, light-paths should be short in order to save bandwidth resources~\cite{Liu201344}. 
The performance of degraded systems is dramatically deteriorated with distance~\cite{Cover:2006:EIT:1146355}. 
A practical example is electrical networks, which suffer from Joule effect, causing power losses. 
We invite the reader to find a rich discussion on diameter-constrained reliability and its applications in~\cite{CEKP2012}.\\

This paper is organized in the following manner. Section~\ref{dcr} formally presents the problem under study, and its computational complexity. 
Section~\ref{exact} shows three exact methods to find network reliability, focused on factorization method. 
The main contributions are included in Section~\ref{factor}. There, the determination of irrelevant links is fully characterized. 
Additionally, we present invariants for the diameter-constrained reliability in a source-terminal scenario, 
which should be considered to find the DCR exactly. Section~\ref{conclusions} presents concluding remarks and trends for future work.

\section{Diameter-Constrained Reliability}\label{dcr}
We will follow the terminology of Michael Ball~\cite{Ball1986}. 
A \emph{stochastic binary system} (SBS) is a tern $(S,\phi,p)$, being $S=\{a_1,\ldots,a_m\}$ a ground-set with $m$ on-off elements, called \emph{components}, 
$\phi:\{0,1\}^m \to \{0,1\}$ a \emph{structure function} that assigns either up ($1$) or down ($0$) to each system state, and $p=(p_1,\ldots,p_m) \in [0,1]^m$ 
a vector that contains \emph{elementary probabilities of operation} for each component. Consider a random vector $X=(X_1,\ldots,X_m)$, 
being $\{X_i\}_{i=1,\ldots,m}$ a set of independent Bernoulli random variables such that $P(X_i=1)=p_i$. The \emph{reliability of an SBS} is the 
number $r$:
\begin{equation}
r = P(\phi(X)=1) = E(\phi(X)) 
\end{equation}
A pathset is a state $x \in \{0,1\}^m$ such that $\phi(x)=1$. A cutset is a state $x \in \{0,1\}^m$ such that $\phi(x)=0$. 
Minimal pathsets (cutsets) are called \emph{minpaths} (\emph{mincuts}). Let us denote $\overline{x}_{i}$ to binary word $x$ 
with the complementary value for bit $i$. A component $i \in S$ is \emph{irrelevant} if $\phi(x)=\phi (\overline{x}_{i})$ for 
all possible states $x \in \{0,1\}^m$. In words, a component is irrelevant when its elementary state does not affect the 
global system state.\\

The classical network reliability problem considers a simple graph $G=(V,E)$ and a set of distinguished nodes $K\subseteq V$, 
called \emph{terminal set}. The corresponding SBS is defined by the following tern:
\begin{itemize}
 \item[1] The ground set is the ordered-set of links: $S=E=(e_1,\ldots,e_m)$. 
 \item[2] A corresponding probability vector: $p=(p_1,\ldots,p_m) \in [0,1]^m$.
 \item[3]  For each $E^{\prime} \subseteq E$,  a binary word $w_{E^{\prime}} = (w_1,\ldots,w_m)$ such that $w_i=1$ if and only if 
$e_i \in E^{\prime}$, and $\phi(w)=1$ if and only if all pair terminals in $G^{\prime}=(V,E^{\prime})$ are connected by some path.
\end{itemize}
The classical network reliability $r$ is historically termed connectedness probability as well~\cite{provan83}. 
In the diameter-constrained scenario, the structure function is modified, and $\phi(w)=1$ additionally requires paths   
with $d$ hops or less between each pair of terminals, being $d$ a positive integer called \emph{diameter}. 
We will denote $R_{K,G}^{d}$ the diameter-constrained reliability.\\

The exact reliability computation is at least as hard as minimum cardinality cutset recognition~\cite{Ball1986}.
Arnon Rosenthal observed that minimum cardinality recognition in the classical reliability problem is precisely Steiner Tree Problem 
in graphs~\cite{Rosenthal}. 
Since this problem is included in Karp's list~\cite{Karp1972}, 
classical reliability computation belongs to the class of $\mathcal{NP}$-Hard problems. 
The reader can observe that the exact diameter-constrained reliability computation is an extension of classical reliability. 
Therefore, it also belongs to the class of $\mathcal{NP}$-Hard problems. H\'ector Cancela and Louis Petingi showed 
the the problem is $\mathcal{NP}$-Hard even in the source terminal case $|K|=2$ when $d\geq 3$; see~\cite{CP2004} for a complete proof. 
A full complexity analysis of different subproblems as a function of $k=|K|$ and $d$ is available in prior works~\cite{CanaleRomero,CP2004,CP2001}. 

\section{Exact Methods}\label{exact}
In order to find the reliability of an SBS, Surech Rai et. al. suggest 
the following classification of exact method~\cite{NET:NET3230250308}:
\begin{itemize}
 \item Inclusion-Exclusion.
 \item Sum of Disjoint Products.
 \item Factorization. 
\end{itemize}

The first one, also called Poincare's formula, is based on a full enumeration of minpaths (or mincuts). 
Assume that $M_1, \ldots,M_l$ is the whole list of minpaths of a certain SBS $(S,\phi,p)$. 
Inclusion-Exclusion method returns the reliability using the following expression:
\begin{equation}\label{ie}
r = P(\bigcup_{i=1}^{l}M_i) = \sum_{i=1}^{l-1}(-1)^{i-1}\sum_{I \subseteq \{1,2,\ldots,l\}; |I|=i} P(\bigcap_{j\in |I|}M_j) 
\end{equation}

The number of minpaths can be exponential with the cardinal of the ground-set $m$. Moreover, the number of terms from Expression~\eqref{ie} 
is even larger. Unless a cancellation of terms or special property is found, a full enumeration of minpaths or mincuts is avoided.\\

Observe that the events $\{M_i\}_{i=1,\ldots,l}$ are non-necessarily disjoint. 
An alternative is to re-write Expression~\eqref{ie}, finding a mutually-exhaustive union of disjoint events. 
Since components fail independently, events are then written as a product of the elementary reliabilities. 
This is the key idea of Sum of Disjoint Products method.\\

Let us have a closer look to the third family of exact methods, called Factorization. The basic idea is to consider 
conditional measure on the operation of some component $i$:
\begin{equation}
r = p_{i} r(\phi|i=1) + (1-p_i)r(\phi|i=0)(1-p_i),
\end{equation}
where $\phi|i=1$ (resp. $\phi|i=0$) is structure $\phi$ conditioned to the event ``component $i$ is in operation'' (resp. failure). 
A shortcoming of this recursive method in its basic form is that it is strongly exponential. 
If the system has irrelevant components they should be discarded, and the process can be largely accelerated. The determination 
of irrelevant components depends on the specific structure under study. 

The first work in Factorization in the field of network reliability is authored by Fred Moskowitz, inspired in electrical networks~\cite{6372698}:
\begin{equation}\label{fact1}
R_{K,G} = p_e R_{K^{\prime},G*e} + (1-p_e)R_{K,G-e},
\end{equation}
being $e \in E$ a certain link of graph $G=(V,E)$ with elementary probability $p_e$, $G-e = (V,E-e)$ the deletion 
graph, $G*e$ the contraction graph (contraction of link $e$) and $K^{\prime}$ is the new terminal-set after link contraction.\\

Since contraction operation is not a diameter invariant, Expression~\eqref{fact1} (sometimes called deletion-contraction formula) 
does not hold for the diameter-constrained reliability. However, a similar expression holds:
\begin{equation}\label{fact2}
R_{K,G}^{d} = p_e R_{K,G^{\prime}} + (1-p_e)R_{K,G-e}, 
\end{equation}
being now $G^{\prime}=G$ but with elementary reliability $p_e=1$. Expression~\eqref{fact2} suggests a recursive solution, 
where links are either perfect or deleted in turns, until a halting condition is met 
(either the network has perfect pathset, or there is no feasible pathset).\\

Recall that a recursive application of Expression~\eqref{fact2} is strongly exponential, and reductions/simplifications to successive 
graphs should be performed. We term \emph{Factorization methods} including these aspects as well. 
In classical network reliability, Factorization theory is mature~\cite{Satyanarayana,doi:10.1137/0214057}. 
However, its extension to the diameter-constrained case deserves further research.

\section{Full Characterization of Irrelevant links in DCR} \label{factor}
H\'ector Cancela et. al. propose a sufficient condition for a link to be irrelevant. 
They state the determination of irrelevant links in a source-terminal context is still an open problem~\cite{CEKP2012}. 

Later effort has been carried-out by Louis Petingi, with a stronger sufficient condition~\cite{petingi2013diameter}. 
A recent analysis shows a third sufficient condition, but it leaves the determination of irrelevant links as an open problem~\cite{CanaleRomeroRubino}.\\

Here, a full characterization of irrelevant links is introduced for a source-terminal scenario first, and later in a $K$-terminal context 
(i.e., for an arbitrary terminal-set $K \subseteq V$). 
First, we will show the three sufficient conditions for the source-terminal case, 
available from the literature presented in a chronological order, 
and the reasons that they fail to recognize irrelevant links in some graphs. Consider an arbitrary graph $G=(V,E)$, 
a two-terminal set $K=\{s,t\}$, a diameter $d$ and a specific link under study $e=\{x,y\}$. 

\begin{figure}[h!]\centering{
	 \begin{tikzpicture}  
	 [scale=1.2,>=stealth,shorten >=0.1pt, auto,  thick,
	nodot/.style={circle, draw = black , very thin, minimum size=12pt, inner sep=0pt, font=\scriptsize },                     
	 ]
	\begin{scope}[xshift=0cm,scale=1]
	\node [nodot] (s) at (0,0) {$s$};
	\node [nodot] (1) at (1,0) {1};
	\node [nodot] (2) at (1,1) {2};
	\node [nodot] (3) at (2,2) {3};
	\node [nodot] (4) at (3,2) {4};
	\node [nodot] (5) at (4,2) {5};
	\node [nodot] (6) at (5,1) {6};
	\node [nodot] (t) at (5,0) {t};

      \draw (s)--(1)--(2)--(3)--(4)--(5)--(6)--(t);
      \draw (1)--(4);      
      \draw (1)--(t);   
	\end{scope}
	\end{tikzpicture}} \caption{Sample graph $G$ with an irrelevant link $e=\{1,2\}$ when $d=6$.}
\label{fig:red}
\end{figure}
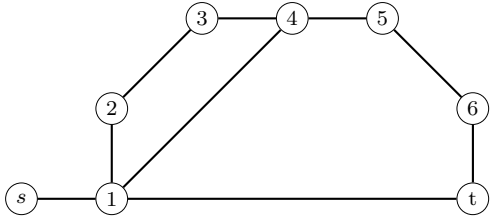

By an elementary analysis, the following conditions are sufficient for link $e$ to be irrelevant:
\begin{itemize}
 \item[1)] $d_G(s,x)+d_G(y,t)\geq d$ and $d_{G}(s,y)+d_G(x,t)\geq d$;
 \item[2)] $d_{G-e}(s,x)+d_{G-e}(y,t)\geq d$ and $d_{G-e}(s,y)+d_{G-e}(x,t)\geq d$;
 \item[3)] $d_{G-y-t}(s,x) + d_{G-s-x}(y,t)\geq d$ and $d_{G-x-t}(s,y) + d_{G-s-y}(x,t)\geq d$. 
\end{itemize}
Let us consider the graph $G$ sketched in Figure~\ref{fig:red}. Observe that link $e=\{1,2\}$ is irrelevant for diameter $d=5$, and even 
for $d=6$ as well. However, $d_G(s,1)+d_G(2,t)=1+2=3<5$, so Condition 1 does not detect that $e$ is irrelevant for $d=5$ nor $d=6$. 
Observe that $d_{G-e}(s,1)+d_{G-e}(2,t)=1+4=5$ and $d_{G-e}(s,2)+d_{G-e}(1,t)=4+1=5$, so Condition $2$ detects that $e$ is irrelevant 
when $d=5$, but it is not the case for $d=6$. Finally, $d_{G-2-t}(s,1)+d_{G-1,s}(2,t)=1+5=6$ and 
$d_{G-1-t}(2,s)+d_{G-s-2}(1,t)=\infty$, so Condition $3$ detects that $e$ is irrelevant in both cases. 
Nevertheless, the reader can check that link $e^{\prime}=\{2,3\}$ is irrelevant when $d=6$, but no sufficient condition detects that 
$e^{\prime}$ is irrelevant.  

A basic result from its definition is that if  a certain link $e$ is irrelevant for a diameter $d$, 
then it will also be irrelevant for any diameter $d^{\prime}\leq d$.\\ 

Now, we will fully characterize irrelevant links. Let $G=(V,E)$ be a simple graph, $K=\{s,t\}$ the terminal set, 
$d$ a positive integer (diameter) and $e=\{x,y\} \in E$ a certain link. Under these conditions, $e$ is relevant 
if and only if there is some $s-t$ path $P$ composed by at most $d$ links, such that $e \in P$.\\

Equivalently, we will find two node-disjoint paths $P_1$ and $P_2$ from nodes $x,y$ to nodes $s,t$ with the minimum length-sum. 
Then, $e$ is irrelevant if and only if path $P= P_1 \cup \{x,y\} \cup P_2$ has more than $d$ links.  
%

In order to find the desired paths $P_1$ and $P_2$, let us extend the original network. 
Consider two artificial nodes $u$ and $z$ with degree $2$. 
Specifically, dode $u$ is connected with terminals $s$ and $t$, while node $z$ is connected with $x$ and $y$. 
We should find two node-disjoint paths $P^{\prime}_1$ 
and $P^{\prime}_2$ between $u$ and $z$ with minimum length-sum. Suurballe's algorithm~\cite{NET:NET3230040204} 
(or Bhandari's algorithm~\cite{DBLP:conf/iscc/Bhandari97}) provides precisely those paths. 
After the deletion of artificial nodes, we obtain the desired paths $P_1$ and $P_2$. We proved the following
\begin{theorem}
Link $e$ is relevant if and only if $l(P_1 \cup P_2) \leq d-1$\\ 
(where the disjoint paths $P_1$ and $P_2$ are found using Suurballe's algorithm). 
\end{theorem}

As a consequence, the deletion of irrelevant links is a DCR invariant. In the most general $K$-terminal context, 
the determination of irrelevant links is performed analogously: just check all pair of terminals whether link $e$ is part of some path 
between two terminals or not. The following elementary operations are DCR invariants as well, 
for a source-terminal configuration:
\begin{itemize}
 \item \emph{Pending-Node}: If the source $s$ (idem terminal $t$) is pending on a link $e=\{s,x\}$ with reliability $p_e$, then we contract link $e$, and 
replace $G$ for its contraction $G*e$. The invariant is $R_{\{s,t\},G}^{d} = R_{\{s^{\prime},t\},G*e}^{d-1}$, being $e^{\prime}$ the 
new source. All non-terminal pending nodes are deleted.
 \item \emph{Perfect-Path}: If a path $P=\{v_1,\ldots,v_n\}$ is an induced subgraph for $G$ and links have elementary reliabilities 
$p_{v_i,v_i+1}$, then we re-assign the link reliabilities $p_{v_i,v_{i+1}}=1$ for all $i=1,\ldots,n-2$ but 
$p_{v_{n-1},v_n}= \prod_{i=1}^{n-1}p_{v_i,v_{i+1}}$.
 \item \emph{Perfect-Neighbors}: if the source $s$ (idem terminal $t$) has all perfect links to its neighbors $N(s)$, then $s \cup N(s)$ is a 
new vertex in $G^{\prime}$ and $R_{\{s,t\},G}^{d} = R_{\{s,t\},G^{\prime}}^{d-1}$.
 \item \emph{Perfect-Cut-Node}: if $v$ is a cut-node (i.e., $G-v$ has more than one component), first delete 
components with all non-terminal nodes (observe that we cannot finish with more than two components). 
Second, apply Perfect-Neighbors to $v$ on both sides.
 \item \emph{Parallel-Links}: If we find two links $e_1$ and $e_2$ from the same nodes with elementary reliabilities $p_{e_1}$ and $p_{e_2}$, 
they are replaced by a new link $e$ with reliability $p_e = p_{e_1}+p_{e_2}-p_{e_1}p_{e_2}$. 
\end{itemize}

These invariants are building blocks of a Factorization method, combined with the deletion of irrelevant links and 
reduction of selected links.

\begin{algorithm}[H]
\caption{$R = Factor(G,s,t,d)$} \label{BLocal}
\begin{algorithmic}[1]
\IF{$HasPefectPath(G,d)$}
\RETURN $R=1$
\ENDIF
\IF{$Distance(s,t)<d$}
\RETURN $R=0$
\ENDIF
\STATE $G \leftarrow Delete(G,Suurballe)$
\STATE $(G,s,t,d) \leftarrow Invariants(G,s,t,d)$
\STATE $e \leftarrow NonPerfectRightMost(E(G))$
\RETURN $(1-p_e) \times Factor(G-e,s,t,d)+p_e \times Factor(G*e,s,t,d)$
\end{algorithmic}
\end{algorithm}

We put all together in $Factor$ Algorithm. It receives the graph $G$, two terminals $s,t$ and a diameter $d$, 
and returns $R=R_{\{s,t\},G}^{d}$. The block of Lines $1$-$6$ test the termination (i.e., either a perfect pathset or no 
feasible pathset). In Line $7$, Suurballe's algorithm is called in order to determine, for each link, whether it is relevant or not. 
Observe that the order of this test does not matter, since the deletion of an irrelevant link does not remove any minpath. 
In Line 8, a list of invariants help to further reduce and simplify the graph. So far, the list has five elementary operations, 
as previously detailed. In Line 9, a certain link is selected in order to perform Factor decomposition. Here, we recommend 
to choose the non-perfect link that is closest to the terminal $t$ (or one of them chosen uniformly at random in case of several 
links). In this way, we improve the activity of Perfect-Neighbors operation (i.e., contracting all nodes close to neighbors from $t$, 
and reducing the diameter in one unit). 

\section{Conclusions and Trends for Future Work}\label{conclusions}
In this paper we discussed exact approaches for the exact diameter-constrained reliability (DCR), 
focused in a source-terminal context. The exact DCR computation belongs to the class of $\mathcal{NP}$-Hard problems, 
since it subsumes the classical network reliability problem. 
Factorization techniques are available for the classical problem. However, the determination of irrelevant links has been 
a shortcoming of previous works in the diameter-constrained version.\\

Here, an efficient method for the determination of irrelevant links is provided for the DCR. Additionally, 
some DCR invariants are included, and a factorization algorithm has been introduced. It greedily selects the closest links 
to one of the terminals, in order to increase the degree of perfect links from the terminals (or reduce the degree after link deletion).\\

Currently, we are implementing this algorithm and similar ones available from the literature in order to perform a faithful comparison 
for both sparse and dense graphs. Additionally, the determination of new DCR invariants is both a challenging and useful task 
in order to develop better exact factorization algorithms for the source-terminal DCR computation.

\bibliography{bib-dcr}
\bibliographystyle{aomalpha}
\end{document}